\documentclass[superscriptaddress,twocolumn,showpacs,prb,floatfix]{revtex4}
\usepackage{graphicx}
\usepackage{tabularx}
\pdfoutput=1
\begin{document}
%\documentclass[aps,twocolumn,showpacs,prb,superscriptaddress]{revtex4}
%\bibliographystyle{apsrevtitle}\prec

%\begin{document}
\title{Extended Scaling for the high dimension and square lattice Ising Ferromagnets}
\author{I.~A.~Campbell}
\affiliation{Laboratoire des Collo\"ides, Verres et Nanomat\'eriaux,
Universit\'e Montpellier II, 34095 Montpellier, France}
\author{P.~Butera}
\affiliation{Istituto Nazionale di Fisica Nucleare, Sezione di Milano-Bicocca, 3 Piazza della Scienza, 20126 Milano, Italy}

\begin{abstract}
In the high dimension (mean field) limit the susceptibility and the second moment correlation length of the Ising ferromagnet depend on temperature as $\chi(T)=\tau^{-1}$ and $\xi(T)=T^{-1/2}\tau^{-1/2}$ exactly over the entire temperature range above the critical temperature $T_c$, with the scaling variable $\tau=(T-T_c)/T$. For finite dimension ferromagnets temperature dependent effective exponents can be defined over all $T$ using the same expressions. For the canonical two dimensional square lattice Ising ferromagnet it is shown that compact "extended scaling" expressions analogous to the high dimensional limit forms give accurate approximations to the true temperature dependencies, again over the entire temperature range from $T_c$ to infinity. Within this approach there is no cross-over temperature in finite dimensions above which mean-field-like behavior sets in.
\end{abstract}

\maketitle
\section{Introduction}
The remarkable critical behavior at second order phase transitions has been intensively studied for many years. In the limit where the temperature $T$ tends to the critical temperature $T_c$, observables $Q(T)$ diverge as
\begin{equation}
Q(T) \sim Q|t|^{-q}
\label{crit}
\end{equation}
where $q$ is the critical exponent and $t$ is the scaling variable; the textbook approach \cite{baxter:82} is to use $t=(T-T_c)/T_c$ as the scaling variable, with small corrections, both analytic and non-analytic, to the strict critical form as soon as $T$ is not infinitesimally close to $T_c$. It is widely considered that there is only a narrow temperature range, the critical region, in the immediate vincinity of $T_c$, where Eqs.~(\ref{crit}) are valid and that they break down beyond this region. Thus the standard protocol for estimating critical exponents from experimental or numerical data is to carry out analyses using Eqs.~(\ref{crit}) with scaling variable $t$, together with finite size scaling (FSS) rules derived using these equations, over as narrow a range of temperature as possible around the critical point, introducing phenomenological corrections to scaling if the data are accurate enough.

In the discussion that follows we will consider the second moment correlation length $\xi_{sm}(T)$ (not the "true" correlation length \cite{fisher:67} $\xi_{true}$) which we will refer to as $\xi(T)$, and the reduced susceptibility $\chi(T)$ (see ~\cite{butera:02} for the definitions). It should be kept in mind that the standard critical observable, the reduced susceptibility $\chi(T)$, is by definition related to the thermodynamic susceptibility $\chi_{th}=(dM/dH)_{H \rightarrow 0}$ through the temperature dependent normalization $\chi(T)=T\chi_{th}(T)$.

From arguments based on high temperature series expansions (HTSE) \cite{butera:02} an "extended scaling" formulation was introduced \cite{campbell:06,campbell:07} in which to leading order the reduced susceptibility is written $\chi(T) \sim [\tau]^{-\gamma}$ and the second moment correlation length is written $\xi(T) \sim T^{-1/2}[\tau]^{-\nu}$, with the scaling variable $\tau = (T-T_c)/T$ \footnote{This scaling variable has already been introduced by many authors, e.g. \cite{fisher:67,fahnle:84,butera:02}}. Here we will discuss two extreme canonical systems : the high dimensional near neighbor Ising ferromagnet on a hypercubic lattice, and the two dimensional Ising ferromagnet on a square lattice. In the high dimensional (mean field) case the critical behavior parameterized this way is strictly exact with temperature independent effective critical exponents over the entire temperature range from $T_c$ to infinity. In the square lattice case the approach leads to compact accurate approximations over the whole range. In the temperature range close to $T_c$ the expression for $\xi(T)$ can be linked to Renormalization Group Theory (RGT) analytic corrections. Using this protocol, experimental or simulation data can be usefully analysed in terms of the critical behavior over a wide temperature range and not only in the very close neighborhood of $T_c$.

\section{High dimension Ising model}

It is instructive to first consider the HTSE for the high dimension Ising ferromagnet in the infinite dimension (or mean field) limit.
For near neighbor interaction spin $1/2$ Ising ferromagnets on [hyper]cubic lattices
the HTSE for the reduced susceptibility \cite{gofman:93} $\chi(\beta)$ and for the second moment of the correlation function $\mu_2(\beta)$ are given exactly by infinite series of terms in powers of $tanh(\beta)$ with $\beta =1/T$ :
\begin{eqnarray}
\chi(\beta) = 1 + (z)tanh(\beta) + (z^2 - z)tanh(\beta)^2 \nonumber\\
+ (z^3 - 2z^2 + z)tanh(\beta)^3 +\cdots
\label{chitanh}
\end{eqnarray}
and
\begin{eqnarray}
\mu_2(\beta) = (z)tanh(\beta) + (2z^2 - z)tanh(\beta)^2 \nonumber\\
+ (3z^3 - 2z^2 +z)tanh(\beta)^3 +\cdots
\label{mu2tanh}
\end{eqnarray}
where $z$ is the number of neighbors ($z= 2d$ for hypercubes in dimension $d$).

The second moment correlation length is defined by
\begin{equation}
\xi(\beta) = [\mu_2(\beta)/z\chi(\beta)]^{1/2}
\label{defxi}
\end{equation}

When $z \rightarrow \infty$ the $(z tanh(\beta))^n$ contribution will dominate at each $n$, and $\beta_c \sim 1/z$ will decrease so $tanh(\beta) \rightarrow \beta$ in all the paramagnetic regime. Hence
\begin{eqnarray}
\chi(\beta) =  1 + z\beta + (z\beta)^2 + (z\beta)^3 + \cdots  \nonumber\\
= ( 1- z\beta )^{-1}
\label{chiinf}
\end{eqnarray}
and
\begin{eqnarray}
\mu_2(\beta) =  z\chi(\beta)\xi(\beta)^2  \nonumber\\
= z\beta[ 1 + 2(z\beta) + 3(z\beta)^2 + 4(z\beta)^3 +\cdots]\nonumber\\
= z\beta( 1- z\beta )^{-2}
\label{mu2inf}
\end{eqnarray}
i.e.
\begin{equation}
\chi(\beta) = (1 - \beta/\beta_c)^{-1}= [(T - T_c)/T]^{-1}
\label{chiinf2}
\end{equation}
and
\begin{equation}
\xi(\beta) = \beta^{1/2}(1 - \beta/\beta_c)^{-1/2}=T^{-1/2}[(T-T_c)/T]^{-1/2}
\label{xiinf}
\end{equation}
exactly for all $\beta$ less than $\beta_c=1/z$, i.e. all $T$ greater than $T_c=z$.

Using as the critical variable $\tau=(1 - \beta/\beta_c)$ instead of $t$ and introducing the prefactor $\beta^{1/2}$ in the expression for $\xi(\beta)$, the critical regime as defined by the scaling expressions :
\begin{equation}
\chi(\beta)=\tau^{-\gamma}
\label{chidef}
\end{equation}
and
\begin{equation}
\xi(\beta)/\beta^{1/2} = \tau^{-\nu}
\label{xidef}
\end{equation}
extends rigorously from $T_c$ to infinite $T$, with
$\beta_c = 1/z$ and temperature independent mean field exponents $\gamma = 1$ and $\nu = 1/2$.

This statement can be reformulated : in the "ideal" mean field limit one can expect correction terms to be inexistant if the critical variable and observables are chosen correctly. The two observables which we have discussed, the reduced susceptibility $\chi(\beta)$ (i.e. $\chi_{th}(\beta)/\beta$) and the "reduced" second moment correlation length $\xi(\beta)/\beta^{1/2}$, show exact critical power law behaviors (as $\tau^{-\gamma}$ and as $\tau^{-\nu}$ respectively) for all $T>T_c$. We can surmise that in finite dimensions there will be correction terms but that the same critical variable and normalized observables will remain the most appropriate for expressing the critical behavior over a wide temperature range.

\section{Square lattice Ising model}

In finite dimensions the extreme simplicity of the mean field case will be lost, but because of the generic structure of the HTSE shown above the general form of equations Eqs.~(\ref{chidef}) and~(\ref{xidef}) (including the non-critical normalization $\beta^{1/2}$ in $\xi(\beta)/\beta^{1/2}$) can be expected to be rather robust \cite{campbell:06}. The exact finite dimension leading critical behaviors for [hyper]cubic ferromagnets when  $\beta \rightarrow \beta_{c}$ can be written
\begin{equation}
\chi(\beta) \rightarrow C_{\chi}(\tau)^{-\gamma}
\label{chicrit}
\end{equation}
and
\begin{equation}
\xi(\beta) \rightarrow [C_{\xi}/\beta_c^{1/2}]\beta^{1/2}(\tau)^{-\nu}
\label{xicrit}
\end{equation}
with critical amplitudes $C_{\chi}$, $C_{\xi}/\beta_c^{1/2}$. For spin $1/2$ the infinite temperature $\beta \rightarrow 0$ values are $\chi(\beta)\rightarrow 1$ and $\xi(\beta) \rightarrow \beta^{1/2}$ for all $d$.
Extended scaling expressions can be written which are analogous to the infinite dimension form, linking the critical limit with the trivial high temperature fixed point limit \cite{campbell:06,campbell:07}. As a first step a minimal modification must be made in order to allow for the fact that in finite dimensions $C_{\chi}$ and $C_{\xi}/\beta_c^{1/2}$ are not equal to exactly $1$. We write :
\begin{equation}
\chi^{*}(\beta) = C_{\chi}[\tau]^{-\gamma}[1 +\tau(1-C_{\chi})/C_{\chi}]
\label{chid}
\end{equation}
and
\begin{equation}
\xi^{*}(\beta) = \beta^{1/2}[(C_{\xi}/\beta_c^{1/2})(\tau)^{-\nu}] [1 +\tau(\beta_c^{1/2}-C_{\xi})/C_{\xi}]
\label{xid}
\end{equation}

These expressions, which depend only on the critical parameters $\beta_c, \gamma, \nu, C_{\chi}$ and $C_{\xi}$, are exact by construction in both the critical and the high temperature limits, and they provide compact approximate expressions for the behavior over the whole range in between. It has been demonstrated that for the standard three dimensional Ising, XY and Heisenberg ferromagnets \cite{campbell:06,campbell:07} extended scaling expressions (defined entirely through the critical parameters appropriate for each particular case) agree with the true $\chi(T)$ and $\xi(T)$ to within better than about $1\%$ over the entire range of temperature from $T_c$ to infinity.

Here we will consider in more detail the particular case of the canonical two dimension square lattice Ising ferromagnet, for which the critical temperature, the critical exponents, and a number of other properties are known exactly from the original work by Onsager and others \cite{baxter:82} and from more recent conformal field theory \cite{belavin:84}. In the square lattice there are no irrelevant operators \cite{fisher:80} except those due to the lattice breaking of rotational symmetry \cite{caselle:01} which can be ignored for present purposes. There are no exact analytic expressions for the susceptibility or the second moment correlation length but many terms of the high temperature series expansions are known \cite{nickel:82,nickel:90,gartenhaus:88}, so $\chi(\beta)$ and $\xi(\beta)$ can be calculated to high precision over the entire range of $\beta$, from $\beta_c$ to zero (i.e. from $T_c$ to infinity).
We have taken advantage of the exact knowledge of the critical
temperature $tanh(\beta_c)=\sqrt2 -1$ (i.e. $\beta_c = 0.440686794\cdots$) and of the critical exponents
$\gamma=7/4$ and $\nu=1$ to form biased Pad\'e approximants of $\chi$ and $\xi^2$,  using up to 48 HTSE coefficients.
Using first-order inhomogeneous differential approximants we would obtain an equivalent accuracy.
Thus $C_{\chi} = 0.962581\cdots$, and $C_{\xi}=0.854221175$. (The ratio of the exact "true" correlation length critical amplitude to the second moment correlation length critical amplitude is $1.000402$ \cite{butera:04}). The extended scaling estimates $\chi^{*}(\beta)$ and $\xi^{*}(\beta)$ (Eqs.~(\ref{chid}) and ~(\ref{xid})) can be written down directly using these values.

Fisher and Burford \cite{fisher:67} some 40 years ago introduced a non-critical factor in the expression for $\xi(\beta)$ for Ising ferromagnets ; they already noted that in the mean field limit the prefactor would be equal to $\beta^{1/2}$ as confirmed by the high dimension limit discussion above. For Ising ferromagnets in dimensions $2$ and $3$ they introduced an "effective range of direct interaction" parameter $r_1(T)$, which they defined through
\begin{equation}
(r_1(T)/\xi(T))^{2-\eta}=1/\chi(T)
\label{r1def}
 \end{equation}
The temperature variation of $r_1(T)$ is mainly due to the non-critical prefactor in the expression for the second moment correlation length $\xi(T)$. Fisher and Burford did not give an explicit expression for $r_1(T)$ in finite dimensions but they calculated it numerically over a wide range of temperature from the HTSE terms known at the time for five different Ising ferromagnets in dimensions 2 and 3 (their Figure 6).
We can recalculate the Fisher-Burford square lattice $r_1(T)$ as defined above from the high precision square lattice Pad\'e approximants values and also from the extended scaling expressions. In Figure 1 we compare over a very wide temperature range
\begin{equation}
r_1(\tau)= \xi_p(\tau)/(\chi_p(\tau))^{1/(2-\eta)}
\label{r1tau}
\end{equation}
obtained directly from the high precision values $\xi_p(\tau), \chi_p(\tau)$ with $r_1^{*}(\tau)$ estimated from the $\xi^{*}(\tau), \chi^{*}(\tau)$ of the extended scaling expressions. On the scale of this plot the two sets of points are almost indistinguishable, showing that the extended scaling expressions are accurate approximations to the exact behavior of the observables.

\begin{figure}
\includegraphics[width=4in]{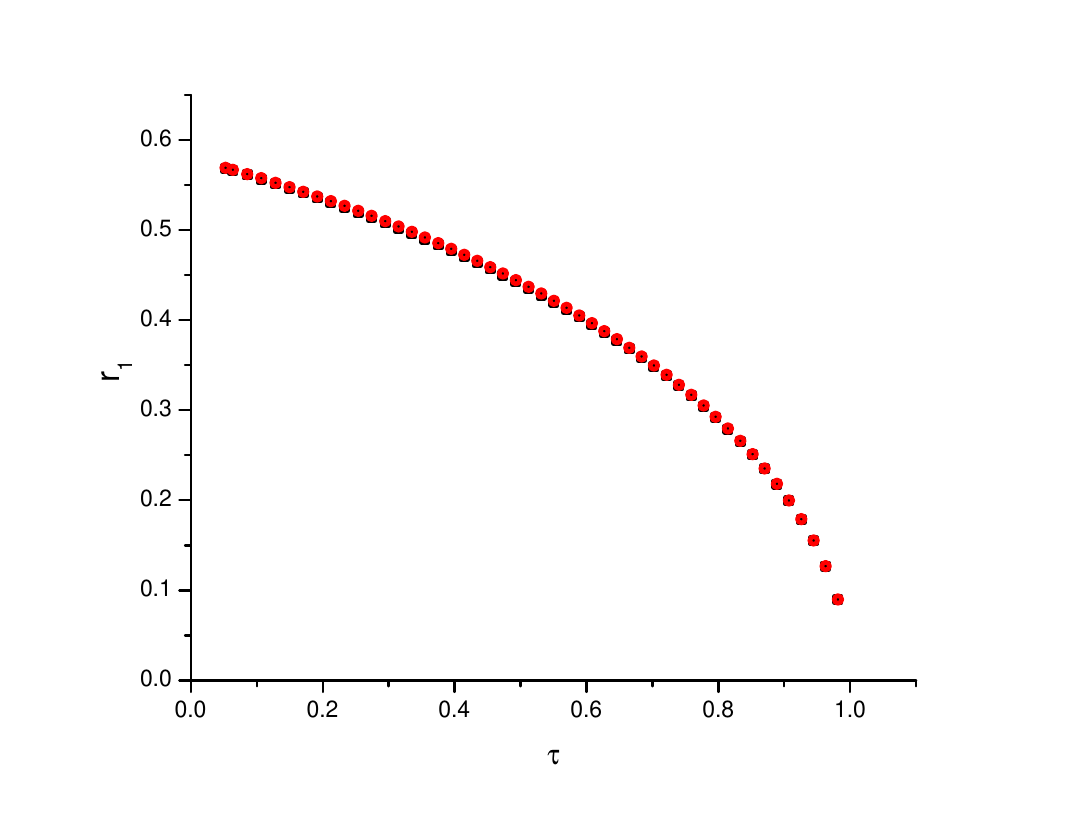}
\caption{The Fisher-Burford "effective range of direct interaction" parameter $r_1(T)$ calculated from the high precision square lattice Pad\'e approximants values (black squares) and from the extended scaling expressions (red circles), as functions of $\tau=1-\beta/\beta_c$.}
\protect\label{fig:1}
\end{figure}

\begin{figure}
\includegraphics[width=4in]{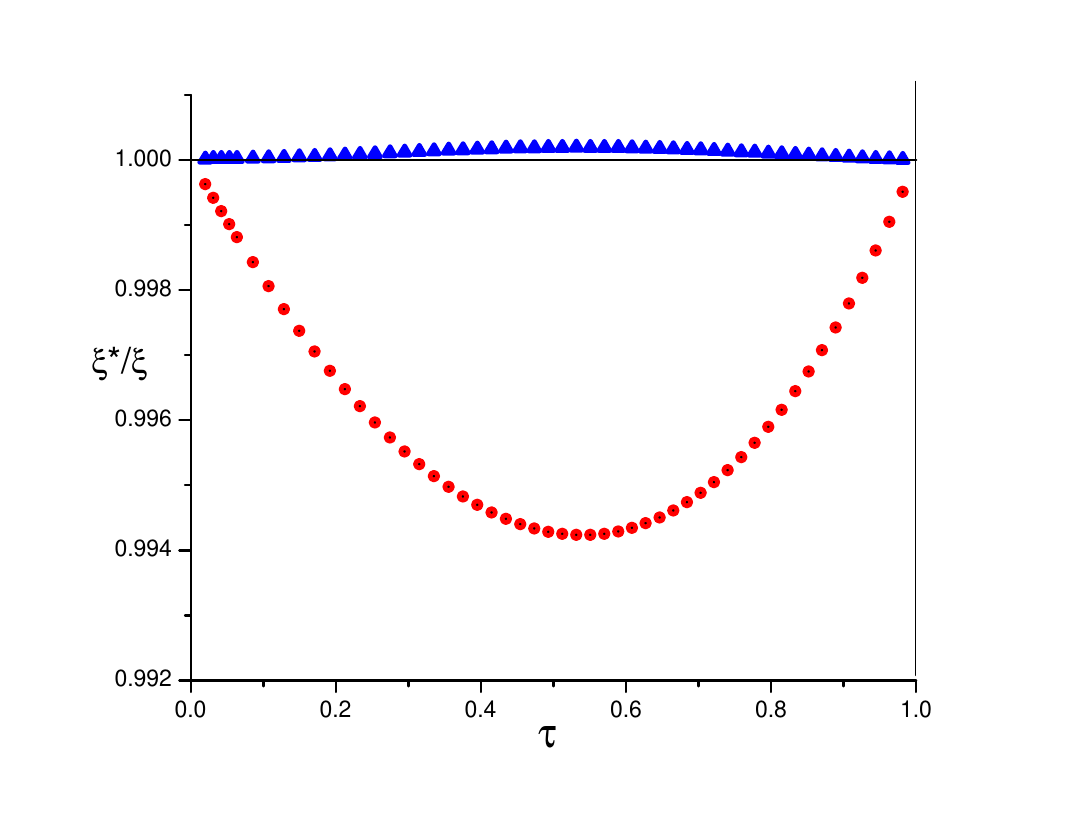}
\caption{The ratio $\xi^{*}(\tau)/\xi_{p}(\tau)$ where the $\xi^{*}(\tau)$ are the extended scaling estimates of the correlation length (red circles one correction term, blue triangles two correction terms) and $\xi_{p}(\tau)$ the high precision data, as a function of $\tau=1-\beta/\beta_c$.}
\protect\label{fig:2}
\end{figure}

\begin{figure}
\includegraphics[width=4in]{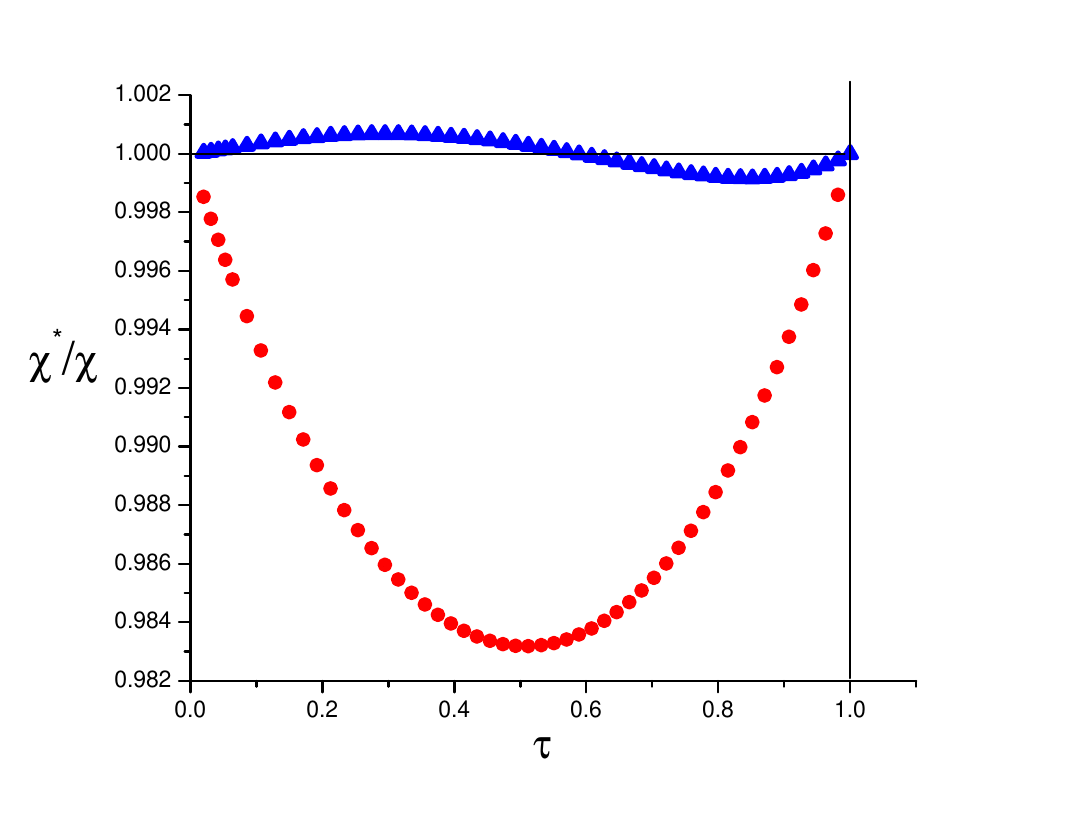}
\caption{The ratio $\chi^{*}(\tau)/\chi_{p}(\tau)$ where the $\chi^{*}(\tau)$ are the extended scaling estimates of the reduced susceptibility (red circles one correction term, blue triangles two correction terms) and $\chi_{p}(\tau)$ the high precision data, as a function of $\tau=1-\beta/\beta_c$.}
\protect\label{fig:3}
\end{figure}

As a more critical test of the extended scaling, in figures 2 and 3 we present the ratios $\xi^{*}(\tau)/\xi_{p}(\tau)$ and $\chi^{*}(\tau)/\chi_{p}(\tau)$ where again $\xi^{*}(\tau)$, $\chi^{*}(\tau)$ are the extended scaling estimates from equations Eq.~(\ref{xid}) and Eq.~(\ref{chid}), and $\xi_{p}(\tau)$, $\chi_{p}(\tau)$ are the high precision calculated values. It can be seen already that without further correction factors the extended scaling values represent the high precision data for $\chi(\tau)$ and for $\xi(\tau)$ to better than $1\%$ over the entire temperature range from $\tau = 0$ to $\tau = 1$. The figures also include $\xi^{*}(\tau)/\xi_{p}(\tau)$ and $\chi^{*}(\tau)/\chi_{p}(\tau)$ ratios where the expressions for $\xi^{*}(\tau)$ and $\chi^{*}(\tau)$ include second correction terms from Eqs.~\ref{gfit} and \ref{chifit}, as will be discussed later.

An alternative manner in which to present the data is to express temperature dependencies of $\chi(\tau)$ and $\xi(\tau)$ in terms of temperature dependent "effective" critical exponents \cite{kouvel:64,fahnle:84,butera:02}. In the spirit of the previous discussion, quite generally for any ferromagnet the temperature dependent effective exponents can be rigorously defined as
\begin{equation}
\gamma_{eff}(\tau) = dlog(\chi(\tau))/dlog(\tau)
\label{gammaeff}
\end{equation}
\begin{equation}
\nu_{eff}(\tau) = dlog(\xi(\beta)/\beta^{1/2})/dlog(\tau)
\label{nueff}
\end{equation}
and
\begin{equation}
\eta_{eff}(\beta) = 2 - dlog(\chi(\beta))/dlog(\xi(\beta)/\beta^{1/2})
\label{etaeff}
\end{equation}
The definition of $\gamma_{eff}(\tau)$ is the same as in references \cite{fahnle:84,butera:02}, but because of the prefactor $\beta^{1/2}$ normalizing $\xi(\beta)$ the other two definitions are not standard. This prefactor is essential to insure sensible high temperature limits in Eq.~\ref{nueff} and Eq.~\ref{etaeff}.
We can note that for the first two parameters an explicit choice must be made for $\beta_c$, while $\eta_{eff}(\beta)$ (obviously linked to the two others) can be calculated from $\chi(\beta)$ and $\xi(\beta)$ data sets without any {\it a priori} knowledge of or estimate for $\beta_c$ as Eq.~\ref{etaeff} does not involve $\tau$. The effective exponents tend to the critical exponent values at $\beta_c$, and from HTSE there are simple exact results for the high temperature limits :  $\gamma_{eff}(0) = 2d\beta_c$, $\nu_{eff}(0)= d\beta_c$ and $\eta_{eff}(0)=0$ in all dimensions for hypercubic lattices.
In the square lattice case $\gamma_{eff}(0) = 1.7627\cdots$ and $\nu_{eff}(0) = 0.8814\cdots$.

Figures 4, 5 and 6 show $\gamma_{eff}(\tau)$, $\nu_{eff}(\tau)$, and $\eta_{eff}(\tau)$, comparing the high precision values from HTSE with the extended scaling values, again with one or two correction terms.

A number of remarks can be made. First, looking only at the high precision data it can be seen that the effective exponents defined through Eqs.~(\ref{gammaeff}),~(\ref{nueff}) and ~(\ref{etaeff}) change smoothly and gradually with temperature over the whole range of temperature from the critical temperature to infinity. For the square lattice $\gamma_{eff}(\tau)$ changes little with temperature while $\nu_{eff}(\tau)$ varies rather more. In the $2d$ case $\eta_{eff}(\tau)$ must necessarily change quite strongly, from the critical value $0.25$ to the infinite temperature value of $0$. In dimension $3$ the absolute value of the change is much weaker.

Secondly, presenting the data in this way is a very sensitive test of the extended scaling expressions; it can be seen that with one correction term agreement with the high precision data is to within better than about $5\%$ for each of the effective exponents over the entire temperature range. With two correction terms the agreement is considerably improved, particularly for $\nu_{eff}$. From the general argument given above, in other ferromagnets one should expect qualitatively similar behavior, with marginal complications due to weak irrelevant operators terms.

\begin{figure}
\includegraphics[width=4in]{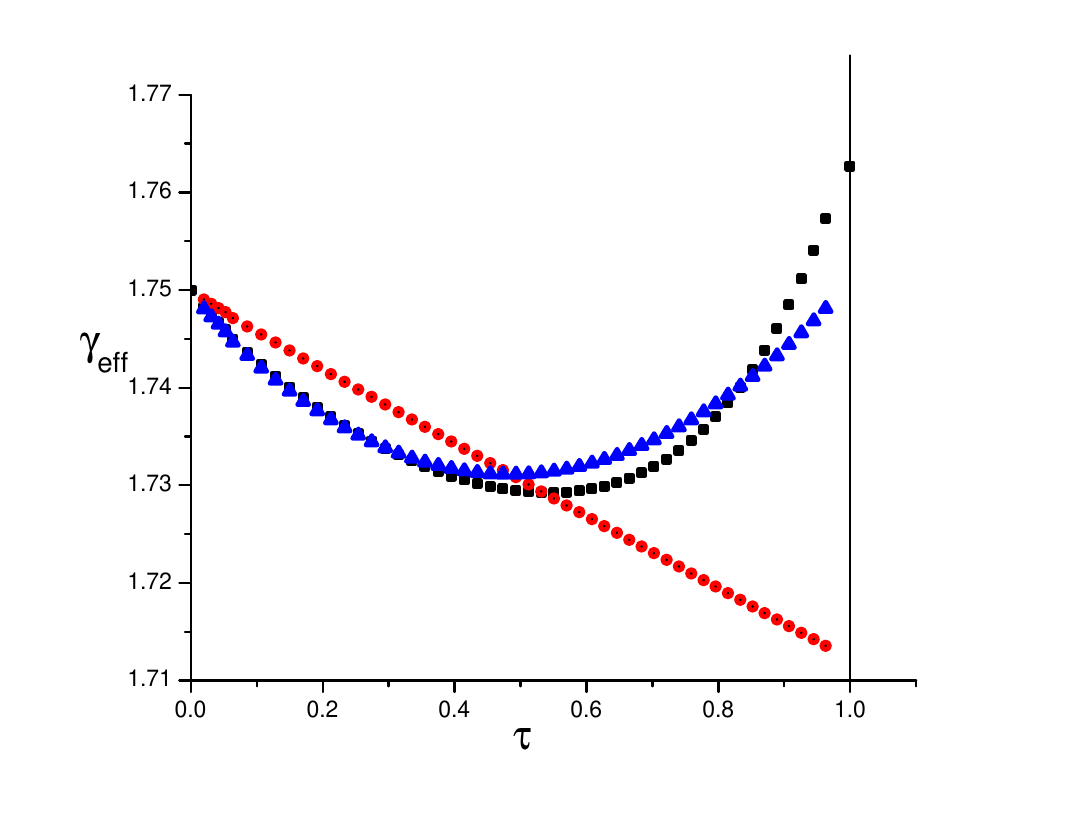}
\caption{The effective exponent $\gamma_{eff}$ calculated from the high precision $\chi_{p}(\tau)$ data (black squares) and from the extended scaling estimates ( red circles one correction term, blue triangles two correction terms), as functions of $\tau=1-\beta/\beta_c$.}
\protect\label{fig:4}
\end{figure}

\begin{figure}
\includegraphics[width=4in]{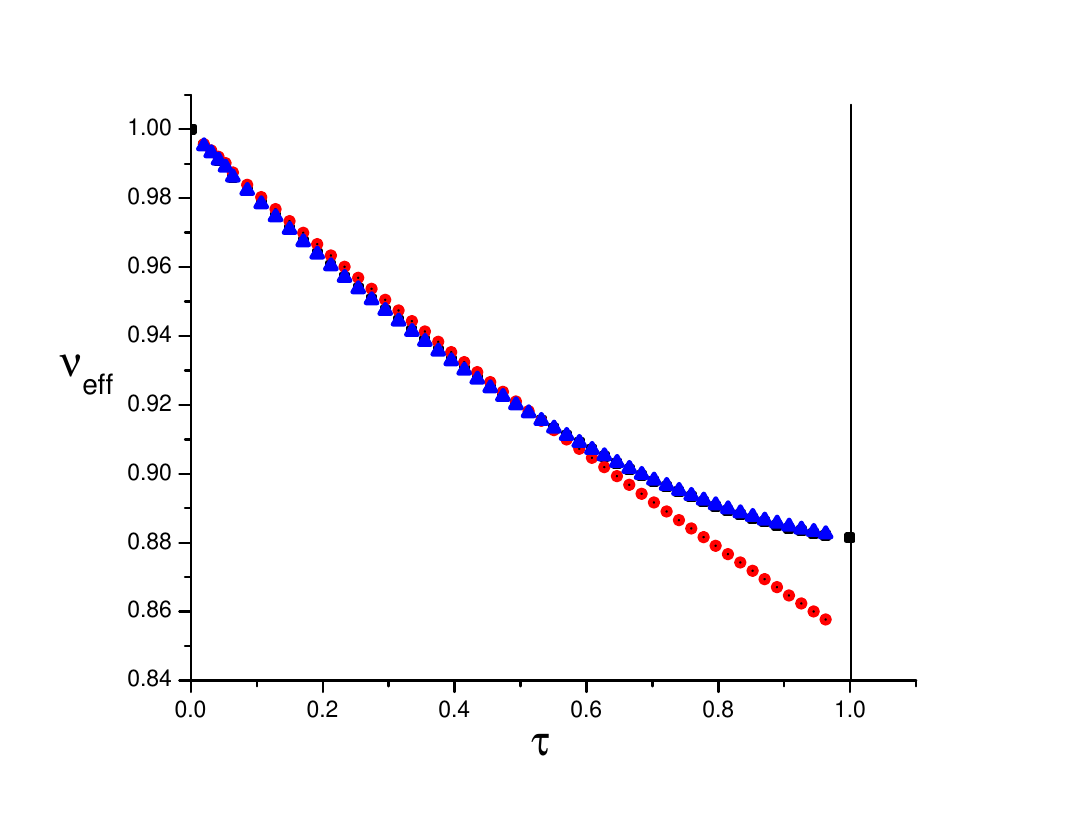}
\caption{The effective exponent $\nu_{eff}$ calculated from the high precision $\xi_{p}(\tau)$ data (black squares) and from the extended scaling estimates ( red circles one correction term, blue triangles two correction terms), as functions of $\tau=1-\beta/\beta_c$.}
\protect\label{fig:5}
\end{figure}

\begin{figure}
\includegraphics[width=4in]{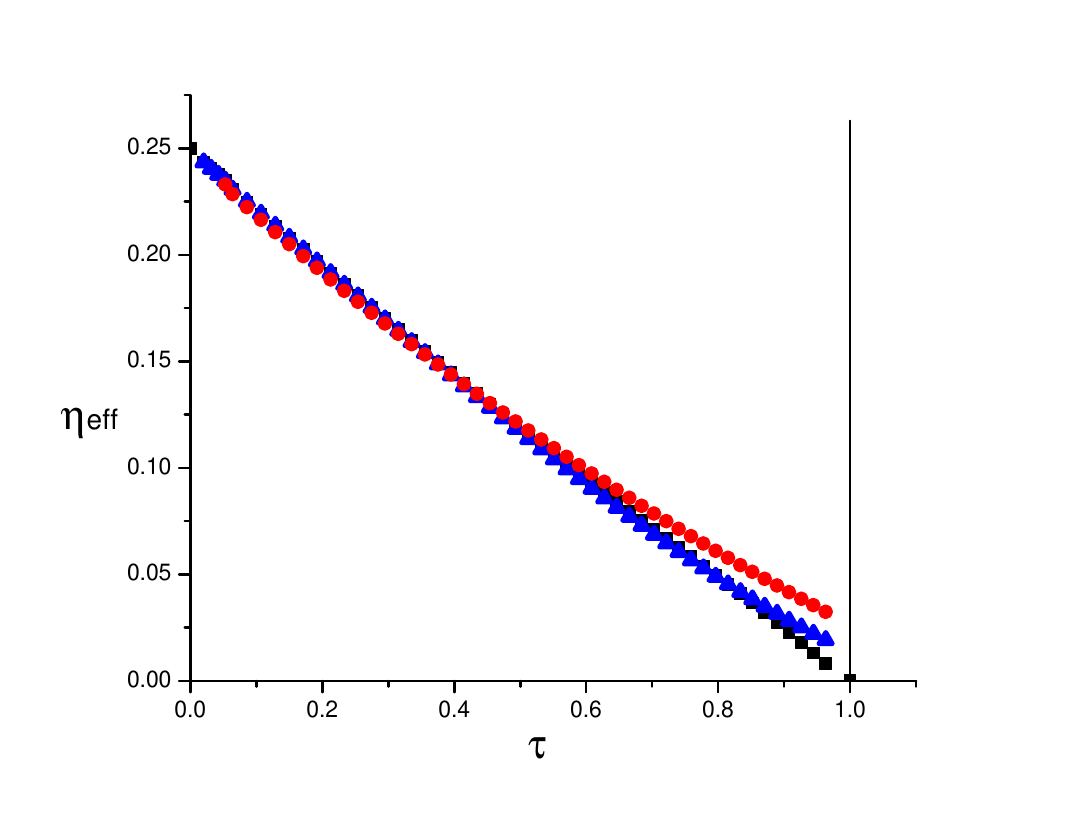}
\caption{The effective exponent $\eta_{eff}$ calculated from the high precision $\chi_{p}(\tau)$ and $\xi_{p}(\tau)$ data (black squares) and from the extended scaling estimates (red circles one correction term, blue triangles two correction terms), as functions of $\tau=1-\beta/\beta_c$.}
\protect\label{fig:6}
\end{figure}

A widely held concept is that one should expect a "cross-over" to mean-field-like effective exponents ($\gamma=1, \nu=1/2, \eta=0$) outside some putative critical region in the neighborhood of $\beta_c$. With the present definitions of the temperature dependent effective exponents there is clearly no such crossover; $\gamma_{eff}$ and $\nu_{eff}$ do not tend to mean-field-like values, although $\eta_{eff}$ does head gradually to the mean field value of $0$.

(Even defining effective exponents in the traditional way using $t$ as the scaling variable and no prefactor would lead to limiting high temperature values $\gamma_{eff}(t_{\infty})=0$, $\nu_{eff}(t_{\infty})=1/2$, and $\eta_{eff}(t_{\infty})=\infty$. $\gamma_{eff}(t_{\infty})$ and $\eta_{eff}(t_{\infty})$ are completely different from the mean field exponents).

\section{Analytic corrections}

The general formalism for corrections to scaling (both analytic and non-analytic) within the Renormalization Group Theory (RGT) is well established (see references \cite{salas:00,caselle:01}).
In the square lattice Ising model for present purposes it can be considered that there are no non-analytic corrections \cite{fisher:80,caselle:01}. This means that one can write for instance the ratio of the inverse of the exact "true" correlation length $1/\xi_{true}(\tau)=ln(coth(\tau))-2\tau$ to the ideal pure critical power law value $\tau/C^{t}_{\xi}$ as an analytic correction factor $g(\tau)$ which is a smooth function having an asymptotic Taylor expansion form, i.e.
\begin{equation}
C^{t}_{\xi}/\xi_{true}(\tau) = \tau g(\tau)=\tau[1+ a_1\tau + a_2\tau^2 + a_3\tau^3 +\cdots]
\label{gdef}
\end{equation}
The angular dependence of the exact leading coefficients of this "large distance 2-point correlation length" analytic correction series has been calculated \cite{salas:00,caselle:01}
 \footnote{It should be noted that Salas and Sokal \cite{salas:00} use a non-standard convention for the Hamiltonan, while Caselle {\it et al} \cite{caselle:01} use a special scaling variable in order to have a compact treatment both above and below $T_c$.}. If $g(\tau)$ is truncated at low order it only gives a useful representation of $\xi_{true}(\tau)$ very close to $T_c$. 

For $\xi$ (it is important to keep in mind that this second moment correlation length is not the "true" correlation length) we choose to define a correction factor $g^{*}(\tau)$ in the extended scaling form :
\begin{equation}
C_{\xi}/\xi(\tau) = ((1-\tau)^{-1/2}\tau^{\nu}g^{*}(\tau)
\label{gstar}
\end{equation}
as $(\beta/\beta_c)^{1/2} =(1 - \tau)^{1/2}$.
Empirical fits can be made to the high precision $\xi_p(\tau)$ square lattice data. We assume a Taylor series for $g^{*}(\tau)$, keeping only two terms (up to order $\tau^{2}$) with the restriction that $g^{*}(1)=C_{\xi}$.  We find that
\begin{equation}
g^{*}(\tau) = 1 - 0.19055\tau + 0.04476\tau^{2}
\label{gfit}
\end{equation}
provides a fit accurate to within less than $0.02\%$ over the entire temperature range from $\tau=0$ to $1$, thus including both the "critical region" and the high temperature region. If the standard $g(\tau)$ correction factor had been used the coefficient for the term in $\tau$ would have become $0.5-0.19055 = 0.30945$. This value is very close to the value of the equivalent coefficient in the $g(\tau)$ series for the "true" correlation length, which is $0.3116$.

The large coefficient of the $\tau$ term in $g^{*}(\tau)$ can be understood as correcting for the difference between $C_{\xi}/\beta_c^{1/2}= 0.854221$ and $1$. The small second coefficient corresponds to a further adjustment which leads to a very significant improvement in the fit. It is remarkable that excellent precision over the entire temperature range is obtained with only two correction terms, validating the use of the $g^{*}(\tau)$ parametrization rather than the $g(\tau)$ form for $\xi_{sm}(\tau)$. We can note that when $g^{*}(\tau)$ is used rather than $g(\tau)$ there are no correction terms to $\xi_{sm}(\tau)$ in the mean field limit.

The square lattice correction terms for $\chi(\tau)$ have been carefully studied \cite{gartenhaus:88}. The coefficients for six leading correction terms are known of which three are exact with additional terms over and above Taylor series terms.
An empirical correction factor for $\chi(\tau)$ with two correction terms gives
\begin{equation}
\chi(\tau) = C_{\chi}\tau^{-7/4}(1+0.0779\tau-0.03903\tau^2)
\label{chifit}
\end{equation}
The leading term in Eq.~\ref{chifit} has a coefficient essentially identical to the exact value ($0.0779032\cdots$) and the phenomenological inclusion of a single further term provides an overall fit up to infinite temperature which is accurate to better than $0.1\%$.

\section{Finite Size Scaling}

 An immediate practical consequence that follows from the discussion above concerns the extraction of critical parameters from numerical studies of systems other than these canonical ferromagnets. If numerical data have been obtained over a wide temperature range and not only in the region very close to $T_c$, direct plots of the effective exponents as defined through Eqs.~(\ref{gammaeff}) to ~(\ref{etaeff}) for the largest sample sizes available can be extrapolated to estimate critical exponents. The temperature dependence of the curves, including the temperature range well above $T_c$, can then give useful indications concerning critical exponents and corrections. Finite size scaling analyses should be made using appropriate expressions derived from the leading extended scaling form. The widely used finite size scaling relation
\begin{equation}
Q(L,T)\sim F[L^{1/\nu}(T-T_c)]
\label{FSS}
\end{equation}
is derived from the Fisher finite size scaling {\it ansatz} $Q(L,T) = F[L/\xi(T)]$ on the assumption that the correlation length behaves as $\xi(T)\sim t^{-\nu}$; as we have seen the latter is always a very poor approximation for $\xi(T)$ except extremely close to $T_c$. Using the Fisher {\it ansatz} together with the extended scaling rule $\xi^{*}(\beta)\sim \beta^{1/2}\tau^{-\nu}$ leads to finite size scaling expressions \cite{campbell:06,campbell:07} which should remain much better approximations over a considerably wider temperature range. If $T_c$ is known to reasonable precision such finite size scaling analyses from numerical data taken over a wide temperature range should give reliable and unbiased estimates for the critical exponents. Ideally an allowance for the correction factor $g^{*}(\tau)$ of the preceding section should also be included in the analysis but this would require very high quality data.

It is important that for each particular system the appropriate extended scaling form should be used. For instance, in spin glasses with symmetrical interaction distributions the right scaling variable is \cite{daboul:04,campbell:06}\\$\tau_{SG}=1-(\beta/\beta_c)^2$.

\section{Conclusion}

One should expect "ideal" critical behavior in the mean field limit ferromagnet, meaning that if the scaling variable and the normalizations of the observables are chosen appropriately all observables should show pure critical power law behavior over the entire temperature range above $T_c$. In the high dimensional limit both the reduced susceptibility $\chi(\tau)$ and the "reduced" second moment correlation length $\xi(\tau)/\beta^{1/2}$ indeed show exact critical power law behaviors (as $\tau^{-\gamma}$ and as $\tau^{-\nu}$ respectively with temperature independent mean field exponents) for all $T>T_c$, validating the use of the scaling variable $\tau=(1-\beta/\beta_c)$ and the normalization for $\xi(\tau)$ through the $\beta^{1/2}$ non-critical prefactor. In other words, in terms of this scaling variable and these observables behavior is "always critical" for the mean field system over the whole temperature range.

For systems in finite dimensions one can no longer expect ideal behavior to be followed exactly at all temperatures, and there will always be deviations from the pure critical power laws as soon as $\tau$ is finite; these deviations can be expressed in terms of correction factors. However in order to reduce the importance of the necessary corrections (and to have sensible high temperature limits) it is judicious to base the choice of scaling variable and scaling expressions on those appropriate for the mean field limit. In agreement with arguments from the general form of the HT series expansions, this leads to extended scaling expressions with a single correction term which are $\chi^{*}(\beta) = C_{\chi}[1-\beta/\beta_c]^{-\gamma}[1 +\tau(1-C_{\chi})/C_{\chi}]$ for the reduced susceptibility and $\xi^{*}(\beta) = \beta^{1/2}[C_{\xi}(1 - \beta/\beta_c)^{-\nu}] [1 +\tau(1-C_{\xi})/C_{\xi}]$ for the second moment correlation length. The adjustment terms have been introduced in order that the expressions tend to their known exact high temperature limits.
For the canonical square lattice Ising ferromagnet with its known critical temperature, exponents, and critical amplitudes $C_{\chi}$ and $C_{\xi}$, the expressions $\chi^{*}(\beta)$ and $\xi^{*}(\beta)$ are compact approximations to the exact behavior which are accurate to within about $1\%$ over the entire temperature range. If the correction series is truncated after only one more term (so that it takes the form $[1+a_{1}\tau+a_{2}\tau^{2}]$) the precision improves to better than $0.1\%$ and better than $0.02\%$ for $\chi(\beta)$ and $\xi(\beta)$ respectively.

The temperature dependence of the observables can be expressed in terms of strictly defined temperature dependent effective exponents, $\gamma_{eff}(\tau)$, $\nu_{eff}(\tau)$ and $\nu_{eff}(\tau)$, which vary smoothly and weakly with temperature. $\gamma_{eff}$ and $\nu_{eff}$ do not tend to mean field values at high temperature. In the square lattice system the two term correction factor for $\xi^{*}(\beta)$ can be compared with the critical analytic correction factor for the "large distance 2-point" correlation length, for which the exact leading terms in the Taylor expansion have been discussed in detail in the RGT formalism \cite{caselle:01}.

Within the extended scaling approach there is no cross-over temperature for finite dimension systems above which mean-field-like behavior sets in.

The present conclusions confirm those already given following analyses of data from other ferromagnets and spin glasses \cite{campbell:06,campbell:07}.
For practical purposes these conclusions should be taken into account when analysing experimental results or when extracting critical parameters from finite size scaling analyses on numerical simulation data.

\end{document}